\documentstyle[11pt,twoside,colloqOHP2010,epsfig]{article}
\begin{document}
\title{Blend Analysis of HATNet Transit Candidates}
\author{J.D. Hartman$^1$, G.\'{A}. Bakos$^1$, G. Torres$^1$}
\affil{$^1$ Harvard-Smithsonian CfA, 60 Garden St., Cambridge, MA, USA [jhartman@cfa.harvard.edu]} 
\begin{abstract}
Candidate transiting planet systems discovered by wide-field
ground-based surveys must go through an intensive follow-up procedure
to distinguish the true transiting planets from the much more common
false positives. Especially pernicious are configurations of three or
more stars which produce radial velocity and light curves that are
similar to those of single stars transited by a planet. In this
contribution we describe the methods used by the HATNet team to reject
these blends, giving a few illustrative examples.
\end{abstract}
\section{Introduction}
To date, the majority of transiting exoplanets (TEPs) have been
discovered by ground-based wide-field surveys such as Super-WASP
(Pollacco et al.~2006) and HATNet (Bakos et al.~2004). For every
planet discovered by these surveys, there is a much greater number of
false positive transit detections (based on statistics to date,
approximately 95\% of the initial HATNet transit detections turn out
to be false positives; see for example Latham et al.~2009). These
false positives typically involve multiple star systems which produce
light curves derived from wide-field survey images that are consistent
with a planet transiting a single star.

While most of the false positives are easily identified and rejected
based on a few high-resolution, low signal-to-noise spectra
(e.g. Latham et al.~2009), or photometric observations that are of
higher precision and spatial resolution than the discovery
observations (e.g. a neighboring eclipsing binary may be blended with
the target on the survey images, but not blended when observed with a
1\,m telescope), some nontrivial false positives produce both radial
velocity (RV) and light curves that are similar to those expected for
a planet transiting a single star. A notable example is the object
OGLE-TR-33, which was selected as a transit candidate by the OGLE-III
survey of the Galactic bulge (Udalski et al.~2002), but shown to be a
hierarchical triple system by Torres et al.~(2004). The blend nature
of this object was revealed by analyzing the spectral line profile
bisector spans to detect subtle variations in the shape of the line
profiles (e.g. Queloz et al.~2001; Torres et al.~2007). Lack of
bisector span (BS) variations at a significant level is now typically
treated as the definitive evidence that a TEP is not a blend.

In this contribution we describe the methods used by the HATNet team
to rule out subtle blend configurations like OGLE-TR-33. We first
describe our {\sc blendanal} program which we use to reject blend
scenarios in cases where a BS analysis is inconclusive (Section 2). We
then discuss several examples from the HATNet survey where a blend
analysis was necessary to determine the nature of the object (Section
3).

\section{The {\sc blendanal} Program}

The {\sc blendanal} program is an implementation of the {\sc blender}
program due to Torres et al.~(2004, 2005, 2010) which pioneered the
approach of fitting a blend model to the photometric observations of
an object. Here we describe the {\sc blendanal} code, highlighting a
few differences from {\sc blender}.

Like {\sc blender}, {\sc blendanal} fits models of one or more stars
to the observations, assuming that one of the stars is eclipsed either
by another star, or by a nonluminous object (planet or brown
dwarf). The properties of the stars are taken from theoretical stellar
isochrones using their masses and assumed ages/metallicities as the
input parameters. The absolute stellar magnitudes are converted into
apparent magnitudes given a distance to each star and a reddening law
along the line of sight. The resulting radii, masses, magnitudes, and
limb darkening coefficients, together with the eclipse ephemeris, the
inclination angle of the eclipsing system and the eccentricity and
argument of periastron are input into {\sc ebop} (Popper \&
Etzel~1981; Etzel~1981; Nelson \& Davis~1972) to derive a model light
curve(s), which is(are) compared to the observed light curve(s). Model
values for the broad-band photometry of the object are also calculated
and compared to observed values. We then compare the $\chi^2$ values
for the best-fit models of various scenarios, including hierarchical
triple systems, blends between a foreground star and a background
eclipsing binary, and binary star systems with one component having a
transiting planet, to the fiducial model of a single star with a
transiting planet, to determine whether the fiducial model is
preferred over all other blend models.

The most important technical differences from {\sc blender} as
described in Torres et al.~(2005) are:
\begin{enumerate}
\item We use the Downhill Simplex Algorithm together with the
  classical linear least squares algorithm to optimize fitted
  parameters. We also include a Markov-Chain Monte Carlo (MCMC)
  routine for optionally determining the uncertainties on these
  parameters, and optionally allow for a grid search over any fitted
  parameter. This method in principle is more likely to find a global
  minimum in a complicated $\chi^2$ landscape than the Torres et
  al.~(2005) version of {\sc blender} which uses a grid search to
  optimize parameters.

\item We include an optional instrumental model to account for
  possible non-physical systematic variations in the light curves. Our
  model is the External Parameter Decorrelation method operated in
  local mode, together with the Trend Filtering Algorithm in global
  mode (see Bakos et al.~2010a). This model is fit to the observations
  simultaneously with the physical model. The differences in the light
  curves between a blended system and a system of a single star with a
  transiting planet may be quite subtle--to robustly distinguish
  between these models, it is important to allow for the possibility
  that some systematic variations in the ground-based light curves
  could be due to instrumental effects. Not doing so overstates the
  confidence with which a blend scenario may be rejected.

\item We allow for a wider range of parameters to be varied, including
  the ages of the stars, their metallicities, and the mass of the
  brightest star in the system. We also allow for different parameter
  choices (such as using the impact parameter of the eclipsing system
  rather than the inclination angle), which leads to important
  differences when conducting an MCMC analysis. In practice, we
  typically fix the mass, age and metallicity of the brightest star to
  reproduce the measured spectroscopic temperature, effective gravity
  and metallicity, but do allow the age and metallicity of a
  background eclipsing binary system to vary.

\item To determine the statistical significance of a $\chi^2$
  difference between two fits we conduct Monte Carlo simulations which
  allow for the possibility of systematic errors that are not
  accounted for by our instrumental model (see Hartman et al.~2009 for
  a description).

\item The model allows for the computation of approximate BS and RV
  values using synthetic spectra together with the {\sc rvsao} package
  (Kurtz \& Mink~1998) for calculating the cross-correlation
  functions. However because the synthetic spectra may differ in
  significant systematic ways from the observed spectra, we do not
  consider these values to be reliable enough to include in the
  $\chi^2$ computation.
\end{enumerate}

\section{Examples from HATNet}

\subsection{Hot Neptunes}

HAT-P-11b is a hot Neptune-mass planet transiting a K4 dwarf star
(Bakos et al.~2010). The orbital semi-amplitude measured with
Keck/HIRES is not large relative to the RV jitter ($K =
11.6$\,m\,s$^{-1}$, jitter$ = 5$\,m\,s$^{-1}$). The BS also exhibit
significant scatter relative to the amplitude of the orbit. As a
result, one cannot definitively rule out blend scenarios based on the
BS. To rule out blend scenarios we made use of an earlier version of
{\sc blendanal} which did not include many of the changes from {\sc
  blender} highlighted in the previous section. In this case it was
not necessary to consider the scenario of a blend between a foreground
star and a background eclipsing binary because the high proper motion
of the star enables the present-day position of the star to be
inspected for contaminating objects using photographic plates from
older sky surveys (Fig.~1). We therefore only treated the case of a
hierarchical triple star system, and found that the light curves could
not be fit unless the two brightest stars were of nearly equal mass,
but in that case the object would have been detected as a double-lined
binary system.

\begin{figure}[ht]
\begin{center}
\epsfig{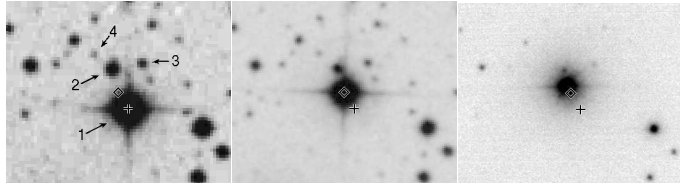}
\caption{The high proper motion ($0\farcs 264$\,yr$^{-1}$) of the
  transiting Neptune host-star HAT-P-11 (labeled as star 1) can be
  used to rule out the possibility that it is a blend with a
  background eclipsing binary. Images of a $2\arcmin \times 1\farcm 7$
  field obtained in 1951, 1989 and 2007 (from left to right) show that
  there is no background object at the present location of HAT-P-11
  that is within 8 magnitudes of HAT-P-11. This figure is taken from
  Bakos et al.~2009.}
\end{center}
\end{figure}

Like HAT-P-11b, HAT-P-26b is a transiting hot Neptune (Hartman et
al.~2010b) for which the BS analysis was inconclusive. Again we
leveraged the high proper motion of HAT-P-26 to rule out a blend with
a background eclipsing binary, and used the version of {\sc blendanal}
described in Section~2 to reject the triple system blend scenario
based on the photometry (Fig.~2).

\begin{figure}[ht]
\begin{center}
\epsfig{width=12cm,file=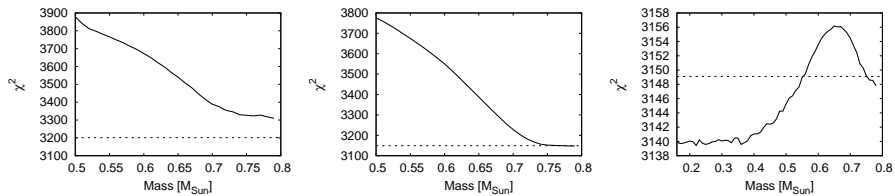}
\caption{Results from using {\sc blendanal} to model the transiting
  Neptune system HAT-P-26 as a hierarchical triple star system (left),
  a binary star where the fainter component has a transiting planet
  (center), and a binary star where the brighter component has a
  transiting planet (right). The mass of the brightest star in each
  case is fixed to $0.79~M_{\odot}$. We show $\chi^2$ as a function of
  the mass of the brighter EB component (left), and the fainter star
  (center and left). In each case points above the dashed line are
  rejected at the $3\sigma$ level. The triple star system is
  completely rejected. Binary star + planet scenarios that fit the
  photometry can be rejected based on the spectra of HAT-P-26.}
\end{center}
\end{figure}

\subsection{Hot Saturns}

The transiting hot Saturns HAT-P-12b (Hartman et al. 2009) and
HAT-P-18b (Hartman et al. 2010a) are two cases where the BS appeared
to correlate with the measured RVs suggesting that these might in fact
be blends. We used {\sc blendanal} to analyze both systems and
concluded that neither could be modelled as a blend. For HAT-P-12 we
used the high proper motion to rule out a blend with a background
eclipsing binary, while for HAT-P-18 this scenario has also been ruled
out with {\sc blendanal}.

To understand the apparent correlation between the BS and RV for
HAT-P-12 we estimated the effect of contamination from scattered
moonlight on the BS measurements and found that much of the BS
variation could be due to this effect. After correcting for the sky
contamination, the BS and RV appeared to be uncorrelated. The tight
relation between the expected BS due to sky contamination and the
measured BS is particularly striking for HAT-P-18 and HAT-P-19
(Fig.~3). 

\begin{figure}[ht]
\begin{center}
\epsfig{width=9cm,file=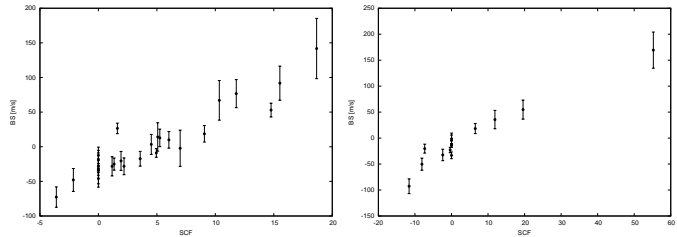}
\caption{Bisector span (BS) vs. the expected BS due to scattered
  moonlight (sky contamination factor; SCF) for HAT-P-18 (left) and
  HAT-P-19 (right). In both cases there is a strong correlation
  between the measured and expected BS.}
\end{center}
\end{figure}

\subsection{HTR294-001}

Our last example is HTR294-001, a transit candidate which exhibits a
striking BS variation that is in anti-phase with the RVs (Fig.~4). In
addition to being in anti-phase, the BS semiamplitude is also an order
of magnitude lower than the RV semiamplitude. Both of these factors
indicate that the transiting object is most likely orbiting the
brightest star in the system. We iteratively use {\sc blendanal} to
model the system as a hierarchical triple, and the TwO-Dimensional
CORrelation ({\sc todcor}; Mazeh \& Zucker~1994) program to extract the RVs,
and find that it is consistent with an F star that is transited by a
very low-mass $0.075$--$0.085~M_{\odot}$ star, which is right at the
stellar--brown dwarf transition, and blended with a K dwarf. The
analysis of this intriguing system continues.

\begin{figure}[ht]
\begin{center}
\epsfig{width=9cm,file=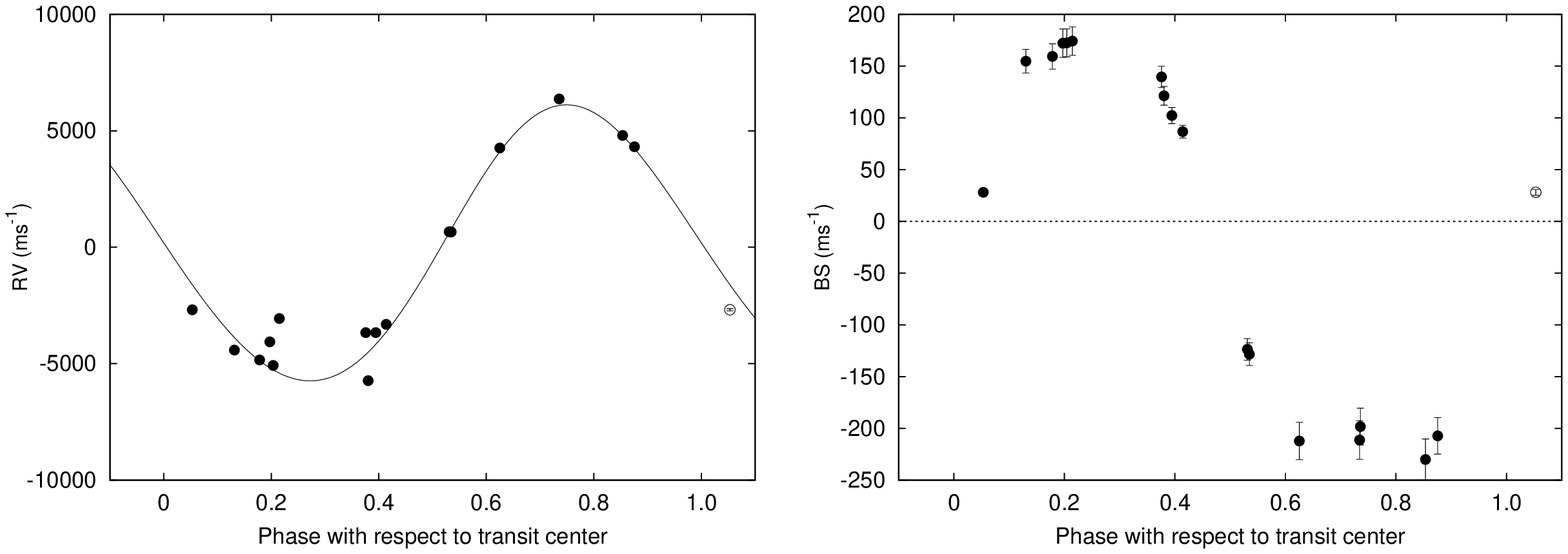}
\caption{Left: RVs for the transit candidate HTR294-001 measured from
  Keck/HIRES spectra obtained with an iodine cell and phased using the
  transit ephemeris. The solid line shows
  the best-fit orbit. Right: bisector spans (BS) for these same
  spectra. The BS are anti-correlated with the RVs and have an
  amplitude that is an order of magnitude lower than the RV
  amplitude. Iteratively applying {\sc blendanal} to model the system
  as a blend, and {\sc todcor} to measure the RVs leads to the conclusion
  that this is a hierarchical triple system.}
\end{center}
\end{figure}

\acknowledgments{HATNet operations have been funded by NASA grants
  NNG04GN74G, NNX08AF23G and SAO IR\&D grants. Work of G.\'A.B. was
  supported by the Postdoctoral Fellowship of the NSF Astronomy and
  Astrophysics Program (AST-0702843).}

\end{document}